# Distributed Estimation and Learning over Heterogeneous Networks

M. Amin Rahimian & Ali Jadbabaie[⋆]

*Abstract*—We consider several estimation and learning problems that networked agents face when making decisions given their uncertainty about an unknown variable. Our methods are designed to efficiently deal with heterogeneity in both size and quality of the observed data, as well as heterogeneity over time (intermittence). The goal of the studied aggregation schemes is to efficiently combine the observed data that is spread over time and across several network nodes, accounting for all the network heterogeneities. Moreover, we require no form of coordination beyond the local neighborhood of every network agent or sensor node. The three problems that we consider are (i) maximum likelihood estimation of the unknown given initial data sets, (ii) learning the true model parameter from streams of data that the agents receive intermittently over time, and (iii) minimum variance estimation of a complete sufficient statistic from several data points that the networked agents collect over time. In each case, we rely on an aggregation scheme to combine the observations of all agents; moreover, when the agents receive streams of data over time, we modify the update rules to accommodate the most recent observations. In every case, we demonstrate the efficiency of our algorithms by proving convergence to the globally efficient estimators given the observations of all agents. We supplement these results by investigating the rate of convergence and providing finite-time performance guarantees.

*Index Terms*—Statistical Learning, Distributed Learning, Distributed Hypothesis Testing, Distributed Detection, Distributed Estimation, Online Learning.

## I. INTRODUCTION

There is a large body of literature on decentralized detection with the notable examples of [1], [2], [3]; and recently there is a renewed interest in this topic due to its applications to sensor and robotic networks [4], [5], [6], [7], [8] and the emergence of new literature considering network of sensor and computational units [9], [10], [11]. Other relevant results investigate the formation and evolution of beliefs in social networks and subsequent shaping of the individual and mass behavior through social learning [12], [13], [14]. The archetype of such models is the one due to DeGroot [15], where agents update their beliefs to a convex combination of their neighbor's beliefs and the coefficients correspond to the level of confidence that each agent puts in each of her neighbors. A variation of this model, where in addition to the neighboring beliefs the agents also receive private signals is considered in [16].

[⋆]Correspondence to: Ali Jadbabaie, Institute for Data, Systems, and Society, Massachusetts Institute of Technology (MIT), Cambridge, MA 02139, USA. (email: jadbabai@mit.edu). This work was supported by ARO MURI W911NF-12-1-0509.

Obtaining a global consensus by combining noisy and unreliable locally sensed data is a key step in many wireless sensor network applications; subsequently, many sensor fusion schemes offer reasonable recipes to address this requirement [17], [18]. In many such applications, each sensor forms an estimate of the field using its local measurements and then the sensors initiate distributed optimization to fuse their local estimates. If all the data from every sensor in the network can be collected in a fusion center, then a jointly optimal decision is readily available by solving the global optimization problem given all the data. However, many practical considerations limit the applicability of such a centralized solution. This gives rise to the distributed sensing problems that include distributed network consensus or agreement [19], [20], [1], and distributed averaging [21]; with close relations to the consensus and coordination problems that are studied in the distributed control theory [22], [23], [24]. Due to the diverse sensing capabilities and other unpredictable physical factors, usually the quality and availability of local observations varries amongst the different sensors and over time. A main focus of this paper is to demonstrate how aggregation schemes can be modified to accommodate the heterogeneity of the sensed data both across time and amongst different sensors.

### A. The Model, Problem Statements and Organization

Consider a set of $n$ agents that are labeled by $[n]$ and interact according to an (undirected) graph $\mathcal{G} = ([n], \mathcal{E})$.[1] The neighborhood of agent $i$ is the set of all agents that it observes and is denoted by $\mathcal{N}_i = \{j \in [n]; (j, i) \in \mathcal{E}\}$; excluding the self-loops: $i \notin \mathcal{N}_i$ for all $i$. We refer to the cardinality of $\mathcal{N}_i$ as the degree of node $i$ and denote it by $d_i$. There is a state $\theta \in \Theta$ that is unknown to the agents and it is chosen arbitrarily by nature from an underlying state space $\Theta$. In the sequel, we pose several estimation and learning problems that the agents face when making decisions given their uncertainty about the true state $\theta$.

In Section II, we consider the case where nodes have an initial data set and they communicate their beliefs in order to combine their local data and come up with the model parameter that best describes all their data collectively; i.e. the global maximum likelihood estimator. In section III, we consider a similar framework where agents instead of starting with an initial data set receive new observations at every round. The data received at every point can provide differing and possibly complementary information about the unknown parameter and the number of data points that is observed at every round varies randomly. In section III, we consider the case where agents observe i.i.d. samples from a distribution and their objective is to estimate the expected value of a complete sufficient statics of the common sampling distribution with as little variance as possible. We consider both the case where they start with initial data state as well as the case where they make new observations at every round. Concluding remarks comparing the nature of update rules and mechanisms of convergence in each case are provided in section V. Proofs and mathematical details can be found in the appendix at the end.

---

[1]Throughout the paper, $\mathbb{R}$ is the set of real numbers, $\mathbb{N}$ denotes the set of all natural numbers, and $\mathbb{N}_0 := \mathbb{N} \cup \{0\}$. For $n \in \mathbb{N}$ a fixed integer the set of integers $\{1, 2, \ldots, n\}$ is denoted by $[n]$, while any other set is represented by a capital Greek or calligraphic letter. Random variables are printed in boldface letters, vectors are represented by lower case Greek or Latin letters with a bar over them and matrices are denoted by upper case Latin letter. We use $\mathbb{1}$ for a column vector of all ones with the proper dimensions and we use $I$ to denote the identity matrix.

In all cases that we consider the mechanisms for convergence rely principally on the fact that increasing powers of a row stochastic matrix behaves in the limit as a rank-one matrix (and $(1/n)\mathbb{1}\mathbb{1}^T$ if the matrix is doubly stochastic). The same principle governs the mixing times of Markov chains and their rate of convergence to the equilibrium distribution, which is determined by the second largest magnitude of the eigenvalues (or the spectral gap) of the coefficients (transition) matrix. These weights can be designed optimally using fastest mixing chains [25], and there are also popular heuristics for assigning the weights, such as the Metropolis-Hastings algorithm. Our choice of weights in this paper is motivated by requirement that the agents construct their update rules based only on their local neighborhood and their sizes (degrees). In the sequel, let the symmetric network graph structure be encoded by its modified adjacency matrix $A = [a_{ij}]_{i,j=1}^n$, defined according to the Metropolis-Hastings weights [25]: $a_{ij} = 1/\max\{d_i, d_j\}$ if $(j,i) \in \mathcal{E}$, and $[A]_{ij} = 0$ otherwise for $i \neq j$; furthermore, $a_{ii} = 1 - \sum_{j \neq i} a_{ij}$.

## II. DISTRIBUTED MAXIMUM LIKELIHOOD ESTIMATION

Suppose that the set of $n$ agents aim to collectively distinguish the true state $\theta$ from a set finitely many possibilities $\Theta$. Each agent $i \in [n]$ has access to a set of $n_i$ initial data points $\mathbf{s}_i^1, \ldots, \mathbf{s}_i^{n_i}$, each of which is identically distributed according to a common distribution $\ell_i(\cdot|\theta)$. In this section we give a procedure so that by forming a belief over the set $\Theta$ and iteratively updating these beliefs, the agents can determine the maximum likelihood estimator of $\theta$ given all the initial data sets: $\{\mathbf{s}_i^1, \ldots, \mathbf{s}_i^{n_i}\}$, $i \in [n]$.

> The agents begin by forming: $\boldsymbol{\gamma}_i(\tilde{\theta}) = \prod_{j=1}^{n_i} \ell_i(\mathbf{s}_i^j|\tilde{\theta})$, and initializing their beliefs to $\boldsymbol{\mu}_{i,0}(\hat{\theta}) = \boldsymbol{\gamma}_i(\hat{\theta})/\sum_{\tilde{\theta} \in \Theta} \boldsymbol{\gamma}_i(\tilde{\theta})$. In any future time period the agents update their belief after communication with their neighboring agents, and according to the following update rule for any $\hat{\theta}$:
> 
> $$\boldsymbol{\mu}_{i,t}(\hat{\theta}) = \frac{\boldsymbol{\mu}_{i,t-1}^{1+a_{ii}}(\hat{\theta}) \prod_{j \in \mathcal{N}_i} \boldsymbol{\mu}_{j,t-1}^{a_{ij}}(\hat{\theta})}{\sum_{\tilde{\theta} \in \Theta} \boldsymbol{\mu}_{i,t-1}^{1+a_{ii}}(\tilde{\theta}) \prod_{j \in \mathcal{N}_i} \boldsymbol{\mu}_{j,t-1}^{a_{ij}}(\tilde{\theta})}. \quad \text{(I)}$$

For any $\hat{\theta} \in \Theta$ we can define $\boldsymbol{\Lambda}(\hat{\theta}) = \sum_{i=1}^n \sum_{j=1}^{n_i} \log(\ell_i(\mathbf{s}_i^j|\hat{\theta}))$, then the global maximum likelihood estimate of $\theta$ given all the initial data points is any member of the set $\Theta^\star := \arg\max_{\hat{\theta} \in \Theta} \boldsymbol{\Lambda}(\hat{\theta})$.

**Theorem 1** (Maximum Likelihood Estimation). *Under* (I), $\lim_{t \to \infty} \boldsymbol{\mu}_{i,t}(\tilde{\theta}) = 1/|\Theta^\star|, \forall \tilde{\theta} \in \Theta^\star$ and $\lim_{t \to \infty} \boldsymbol{\mu}_{i,t}(\tilde{\theta}) = 0, \forall \tilde{\theta} \notin \Theta^\star$ *almost surely, for all* $i \in [n]$. *In particular, if* $\Theta^\star = \{\theta^\star\}$ *is a singleton, then* $\lim_{t \to \infty} \boldsymbol{\mu}_{i,t}(\theta^\star) = 1$, *almost surely for all* $i$. *Hence, after a large enough number of iterations any agent* $i \in [n]$ *can recover* $\theta^\star$ *as* $\theta^\star = \arg\max_{\tilde{\theta} \in \Theta} \mu_{i,t}(\tilde{\theta})$.

In Appendix A, where Theorem 1 is proved we also give a more detailed description of the belief convergence result claimed here; in particular, we characterize a finite time **T**, such that any agent $i \in [n]$ can recover $\theta^\star$ by $\theta^\star = \arg\max_{\tilde{\theta} \in \Theta} \mu_{i,t}(\tilde{\theta})$ at all $t > \mathbf{T}$, cf. (13) of the appendix and the explanations therein.

It is instructive to consider the following unweighted version of (I):

$$\boldsymbol{\mu}_{i,t}(\hat{\theta}) = \frac{\boldsymbol{\mu}_{i,t-1}(\hat{\theta}) \left( \prod_{j \in \mathcal{N}_i} \frac{\boldsymbol{\mu}_{j,t-1}(\hat{\theta})}{\nu_j(\hat{\theta})} \right)}{\sum_{\tilde{\theta} \in \Theta} \boldsymbol{\mu}_{i,t-1}(\tilde{\theta}) \left( \prod_{j \in \mathcal{N}_i} \frac{\boldsymbol{\mu}_{j,t-1}(\tilde{\theta})}{\nu_j(\tilde{\theta})} \right)}. \quad (1)$$

This update can be derived as a Bayesian heuristic that a group member uses to update her belief after listening to the beliefs of her neighbors [26], [27]. Following the Bayesian heuristic belief updates in (1), the asymptotic beliefs will be uniformly supported over $\tilde{\theta} \in \Theta^\star$, where $\Theta^\diamond := \arg\max_{\tilde{\theta} \in \Theta} \sum_{i=1}^n \alpha_i \log(\ell_i(\mathbf{s}_i|\tilde{\theta}))$. In particular, if the sum of signal log-likelihoods weighted by node centralities is uniquely maximized by $\theta^\diamond$, i.e. $\{\theta^\diamond\} = \Theta^\diamond$, then $\lim_{t \to \infty} \boldsymbol{\mu}_{i,t}(\theta^\diamond) = 1$ almost surely for all $i \in [n]$. The fact that log-likelihoods in $\Theta^\diamond$ are weighted by the node centralities is a source of inefficiency for the asymptotic outcome of the group decision process. This inefficiency is warded off in especially symmetric typologies, where in and out degrees of all nodes in the network are the same. In these so-called balanced regular digraphs, there is a fixed integer $d$ such that all agents receive reports from exactly $d$ agents, and also send their reports to some other $d$ agents; $d$-regular graphs are a special case, since all links are bidirectional and each agent sends her reports to and receive reports from the same $d$ agents. In such structures $\overline{\alpha} = (1/n)\mathbb{1}$ so that $\Theta^\star = \Theta^\diamond$ and the support of the consensus belief identifies the global maximum likelihood estimator (MLE); i.e. the maximum likelihood estimator of the unknown $\theta$, given the entire set of observations from all agents in the network. In fact, the heuristic group decision outcome demonstrates also a second form of departure from optimality in that he agents effectively become certain about the truth state of $\theta^\diamond$, in spite of their essentially bounded aggregate information and in contrast with the rational (optimal) belief $\boldsymbol{\mu}^\star$ that is given by the Bayes rule:

$$\boldsymbol{\mu}^\star(\hat{\theta}) = \frac{\prod_{j \in \mathcal{N}_i} \ell_j(\mathbf{s}_j|\hat{\theta})}{\sum_{\tilde{\theta} \in \Theta} \prod_{j \in \mathcal{N}_i} \ell_j(\mathbf{s}_j|\tilde{\theta})},$$

and do not discredit or reject any of the less probable states.

III. LEARNING FROM INTERMITTENT STREAMS OF DATA

In this section, we consider a network of agents that make streams of observations intermittently and communicate their beliefs at every time period. At any time $t$, agent $i$ makes $\mathbf{n}_{i,t}$ i.i.d. observations $\mathbf{s}_{i,t}^1, \ldots, \mathbf{s}_{i,t}^{\mathbf{n}_{i,t}}$ that are distributed according to $\ell(\cdot|\theta)$; and the numbers of observations at each time period: $\{\mathbf{n}_{i,t}, t \in \mathbb{N}\}$ constitute a sequence of i.i.d. signals with mean $\mathbb{E}\{\mathbf{n}_{i,t}\} = \nu_i$. The agents aim to determine the true state $\theta$ from their stream of observations.

> Every time $t \in \mathbb{N}_0$, each agent forms the likelihood product of the signals that it has received at that time-period: $\boldsymbol{\gamma}_{i,t}(\tilde{\theta}) = \prod_{j=1}^{n_{i,t}} \ell_i(\mathbf{s}_{i,t}^j|\tilde{\theta})$, if $n_{i,t} \geq 1$, and $\boldsymbol{\gamma}_{i,t}(\tilde{\theta}) = 1$ if $n_{i,t} = 0$. It then updates its belief according to:
>
> $$\boldsymbol{\mu}_{i,t}(\hat{\theta}) = \frac{\boldsymbol{\gamma}_{i,t}(\hat{\theta}) \boldsymbol{\mu}_{i,t-1}^{a_{ii}}(\hat{\theta}) \prod_{j \in \mathcal{N}_i} \boldsymbol{\mu}_{j,t-1}^{a_{ij}}(\hat{\theta})}{\sum_{\tilde{\theta} \in \Theta} \boldsymbol{\gamma}_{i,t}(\tilde{\theta}) \boldsymbol{\mu}_{i,t-1}^{a_{ii}}(\tilde{\theta}) \prod_{j \in \mathcal{N}_i} \boldsymbol{\mu}_{j,t-1}^{a_{ij}}(\tilde{\theta})}, \quad \text{(II)}$$
>
> initialized by: $\boldsymbol{\mu}_{i,0}(\hat{\theta}) = \boldsymbol{\gamma}_{i,0}(\hat{\theta}) / \sum_{\tilde{\theta} \in \Theta} \boldsymbol{\gamma}_{i,0}(\tilde{\theta})$.

**Theorem 2** (Learning from Intermittent Streams). *Let $\Lambda_i(\hat{\theta}, \check{\theta}) = \mathbb{E}_\theta\{\log(\ell_i(\mathbf{s}_{i,0}|\hat{\theta})/\ell_i(\mathbf{s}_{i,0}|\check{\theta}))\}$ for all $i \in [n]$, and any pair of states $\hat{\theta}, \check{\theta} \in \Theta$. If $\sum_{i=1}^n \nu_i \Lambda_i(\hat{\theta}, \theta) < 0$ for all $\hat{\theta} \neq \theta$, then under (II), $\lim_{t \to \infty} \boldsymbol{\mu}_{i,t}(\theta) = 1$, for all $i$. Moreover, the learning is asymptotically exponentially fast with the rate equal to $\min_{\hat{\theta} \neq \theta}\{(-1/n) \sum_{i=1}^n \nu_i \Lambda_i(\hat{\theta}, \theta)\}$.*

To understand the nature of the convergence result and learning rate in Theorem 2, consider the special case where each agent at every time $t$ may or may not have access to a sample point $\mathbf{s}_{i,t}$ and the accessibility of the new measurement $\mathbf{s}_{i,t}$ is determined by the outcome of an idependent coin flip with success probability $p_i$, i.e. $\{\mathbf{n}_{i,t}, t \in \mathbb{N}\}$ are i.i.d $Bernoulli(p_i)$ variables. Then the convergence rate in Theorem 2 becomes $\min_{\hat{\theta} \neq \theta}\{(-1/n) \sum_{i=1}^n p_i \Lambda_i(\hat{\theta}, \theta)\}$, which decreases linearly with the decreasing probability of making new obsrevations. Also note that $\Lambda_i(\hat{\theta}, \theta) = \mathbb{E}_\theta\{\log(\ell_i(\mathbf{s}_{i,0}|\hat{\theta})/\ell_i(\mathbf{s}_{i,0}|\theta))\} := -D_{KL}\left(\ell_i(\cdot|\hat{\theta})||\ell_i(\cdot|\theta)\right) \leqslant 0$, where $D_{KL}(\cdot||\cdot) \geq 0$ is the Kullback-Leibler divergence. It measures a psudo-distance between the two distributions and it is strictly positive whenever $\ell_i(\cdot|\hat{\theta}) \not\equiv \ell_i(\cdot|\theta)$, i.e. the two distributions disagree over a non-trivial (nonzero measure) set [28, Theorem 2.6.3]; hence, the closer the alternative distributions are to the true distributions the slower is the rate.

IV. MINIMUM VARIANCE UNBIASED ESTIMATION AND ONLINE LEARNING OF THE EXPECTED VALUES

In this section, we allow the parameter space $\Theta$ to be any measurable set, and in particular not necessarily finite. Consider again the network of $n$ agents and suppose that each agent $i \in [n]$ observes an i.i.d. samples $\mathbf{s}_i$ from a common distribution $\ell(\cdot|\theta)$ over a measurable sample space $\mathcal{S}$. We assume that $\ell(\cdot|\theta)$ belongs to a one-parameter exponential family so that it admits a probability density or mass function that can be expressed as

$$\ell(s|\theta) = \tau(s) e^{\alpha(\theta)^T \xi(s) - \beta(\alpha(\theta))}, \quad (2)$$

where $\xi(s) \in \mathbb{R}$ is a measurable function acting as a complete sufficient statistic for the i.i.d. random samples $\mathbf{s}_i$, and $\alpha : \Theta \to \mathbb{R}$ is a mapping from the parameter space $\Theta$ to the real line $\mathbb{R}$, $\tau(s) > 0$ is a positive weighting function, and

$$\beta(\alpha) := \ln \int_{s \in \mathcal{S}} \tau(s) e^{\alpha \xi(s)} ds, \quad (3)$$

is a normalization factor known as the log-partition function. In (2), $\xi(\cdot)$ is a complete sufficient statistic for $\theta$. It is further true that $\sum_{i=1}^n \xi(\mathbf{s}_i)$ is a complete sufficient statistic for the $n$ i.i.d. signals that the agents have received [29, Section 1.6.1]. The agents aim to estimate the expected value of $\xi(\cdot)$: $m_\theta = \mathbb{E}\{\xi(\mathbf{s}_i)\}$, with as little variance as possible. The Lehmann-Scheffé theory (cf. [30, Theorem 7.5.1]) implies that any function of the complete sufficient statistic that is unbiased for $m_\theta$ is the almost surely unique minimum variance unbiased estimator of $m_\theta$. In particular, the minimum variance unbiased estimator of $m_\theta$ given the initial data sets of all nodes in the network is given by: $\mathbf{m}_n = (1/n) \sum_{i=1}^n \xi(\mathbf{s}_i)$. The agents can compute this value using any average consensus algorithm [31]; guaranteeing convergence to average of the initial values asymptotically.

> The agents initialize with: $\boldsymbol{\mu}_{i,0} = \xi(\mathbf{s}_i)$, and in any future time period the agents communicate their values and update them according to the following rule:
>
> $$\boldsymbol{\mu}_{i,t} = a_{ii} \boldsymbol{\mu}_{i,t-1} + \sum_{j \in \mathcal{N}_i} a_{ij} \boldsymbol{\mu}_{j,t-1}. \quad \text{(III)}$$

The mechanisms for convergence in this case rely on the product of stochastic matrices, similar to mixing of Markov chains (cf. [32], [9]); hence, many available results on mixing rates of Markov chains can be employed to provide finite time grantees after $T$ iteration of the average consensus algorithm for fixed $T$. Such results often rely on the eigenstructure (eigenvalues/eigenvectors) of the communication matrix $A$, and the facts that it is a primitive matrix and its ordered eigenvalues satisfy $-1 < \lambda_n(A) \leq \lambda_{n-1}(A) \leq \ldots \leq \lambda_1(A) = 1$, as a consequence of the Perron-Frobenius theory [33, Theorems 1.5 and 1.7].

**Theorem 3** (Minimum Variance Unbiased Estimation). *Under* (III), $\lim_{t\to\infty} \boldsymbol{\mu}_{i,t} = \mathbf{m}_n$ *almost surely, for all $i$. Furthermore,* $|\boldsymbol{\mu}_{i,t} - \mathbf{m}_n| \leq \epsilon$, *whenever*
$$t > \left(\log(\epsilon) - \log(\mathbf{M}_n \sqrt{n-1})\right) / \log \beta^\star,$$
*where* $\mathbf{M}_n = \max_{i \in [n]} |\xi(\mathbf{s}_i)|$ *and* $\beta^\star = \max\{\lambda_2(A), |\lambda_n(A)|\}$.

We now take a brief look at the case where the initial data sets are of different sizes $n_i$, so that each agent has access to a set of $n_i$ initial data points $\mathbf{s}_i^1, \ldots, \mathbf{s}_i^{n_i}$, each of which is identically distributed according to the common exponential family distribution $\ell(\cdot|\theta)$. We explain how (III) should be modified to accommodate the varying sample sizes. In this case, the globally efficient (minimum variance) estimator of the mean sufficient statistic $m_\theta$ given all the initial data sets is as follows: $\mathbf{m}_n^\star = \left(1/\sum_{p=1}^n n_p\right) \sum_{i=1}^n \sum_{j=1}^{n_i} \xi(\mathbf{s}_i^j)$. The agents can initialize with: $\boldsymbol{\mu}_{i,0} = (1/n_i) \sum_{j=1}^{n_i} \xi(\mathbf{s}_i^j)$ for all $i$; however, to ensure convergence to the right limit the coefficients of the linear update rule in (III) should be modified in accordance with the initial sample sizes. Let $\delta_i = n_i/\sum_{p=1}^n n_p$; the following modification of the Metropolis - Hastings weights explained in [25], [34], incorporates the sample sizes and ensures convergence of the linear iterations in (III) to the right limit $\mathbf{m}_n^\star$:

$$a_{ij} = \begin{cases} \frac{1}{d_i} \min\{1, \frac{\delta_j d_i}{d_j \delta_i}\} & \text{if } (j,i) \in \mathcal{E}, \\ 1 - \sum_{j \neq i} a_{ij} & \text{if } i = j, \\ 0 & \text{otherwise.} \end{cases} \quad (4)$$

Next, suppose that every time $t \in \mathbb{N}$, each agent $i \in [n]$ receives an i.i.d. sample $\mathbf{s}_{i,t}$, in addition to communicating their current estimates $\boldsymbol{\mu}_{i,t}$. All signals $\{\mathbf{s}_{i,t} : i \in [n], t \in \mathbb{N}\}$ are distributed according to the same distribution $\ell(\cdot|\theta)$, and as before the agents aim to estimate the expected value of the complete sufficient sufficient statistic $\xi(\cdot)$ with as little variance as possible. Here we propose a $1/t$ discounting of new samples with increasing time $t$. This would enable the agents to learn the true value $m_\theta$ asymptotically almost surely; and in such a way that the variance of their estimates decreases as $1/t$: linearly in time. The exact upper bound for Var$\{\boldsymbol{\mu}_{i,t}\}$ is derived in in Appendix D as follows:

$$\text{Var}\{\boldsymbol{\mu}_{i,t}\} \leq \frac{n(n-1)\mathbb{E}\{\xi(\mathbf{s}_{j,\tau})^2\}}{t(1-\beta^{\star 2})} + \frac{\text{Var}\{\xi(\mathbf{s}_{1,1})\}}{nt}. \quad (5)$$

We can further use the properties of the exponential family to express the expectation and variance of the complete sufficient statistic $\xi(\cdot)$ in terms of the first and second derivatives of the log-partition function given in (3), cf. [29, Theorem 1.6.2]: $\mathbb{E}\{\xi(\mathbf{s}_{j,\tau})\} = \beta'(\alpha(\theta))$ and Var$\{\xi(\mathbf{s}_{1,1})\} = \beta''(\alpha(\theta))$. Hence, (5) becomes:

$$\text{Var}\{\boldsymbol{\mu}_{i,t}\} \leq \frac{n(n-1)(\beta'(\alpha(\theta))^2 + \beta''(\alpha(\theta)))}{t(1-\beta^{\star 2})} + \frac{\beta''(\alpha(\theta))}{nt}, \quad (6)$$

The preceding upper-bound can be used to provide finite-time guarantees for the quality of the estimate $\boldsymbol{\mu}_{i,T}$ at any node $i$ and after a finite termination time $T$. These bounds are comprised of two additive terms: the first terms on right-hand sides of (5) and (6) capture the rate at which the powers of Metropolis-Hastings weight matrix $A$ approach their limit: $A^t \to \frac{1}{n} \mathbf{1}\mathbf{1}^T$ as $t \to \infty$; the second term captures the diminishing variance of the estimates with the increasing number of samples, as gathered by all the agents in the network. The latter is a simple consequence of the Chebyshev inequality applied to the entire set of $nt$ samples that are gathered by all the $n$ agents up to time $t$. On the other hand, the first term on the right-hand side of the bounds is governed by the mixing rate of the Metropolis-Hastings weights; in particular, it is influenced by the structure of the network through $\beta^\star$: the second largest magnitude of the eigenvalues of matrix $A$. The same structural effect appears through $\beta^\star$ in the bound claimed in Theorem 3. A similar effect can be observed through $\alpha^\star$ from the expression of the finite-time **T** in the proof of Theorem 1, (Appendix A).

---

Initializing $\boldsymbol{\mu}_{i,0}$ arbitrarily, in any future time period $t \geq 1$ the agents observe a signal $\mathbf{s}_{i,t}$, communicate their current values $\mu_{i,t-1}$, and update their beliefs to $\mu_{i,t}$, according to the following rule:

$$\boldsymbol{\mu}_{i,t} = \frac{t-1}{t}\left(a_{ii}\boldsymbol{\mu}_{i,t-1} + \sum_{j \in \mathcal{N}_i} a_{ij}\boldsymbol{\mu}_{j,t-1}\right) + \frac{1}{t}\xi(\mathbf{s}_{i,t}). \quad \text{(IV)}$$

---

**Theorem 4** (Online Learning of Expected Values). *Under* (IV), $\lim_{t\to\infty} \boldsymbol{\mu}_{i,t} = m_\theta$ *almost surely, for all $i$. Furthermore, Var$\{\boldsymbol{\mu}_{i,t}\} = O(1/t)$ and $\mathbb{E}\{\mu_{i,t}\} = m_\theta$ for all $t$.*

Unlike the log-linear update rules which could be easily modified to accommodate intermittent data streams with varying sizes (compare (I) and (II)), the linear update rules are not amenable to heterogeneities in the network. It is due the requirement to discount the new observations with increasing time and the need to adapt the linearity coefficients to the varying sample sizes (see (4)). These factors make the linear update rules unnameable to the case of intermittent observations (compare (III) and (IV)).

## V. CONCLUDING REMARKS

The log-linear structure of the proposed belief update rules in (I) and (II) is motivated by our earlier results on the Bayesian without Recall (BWR) model of learning and inference over networks [35]. Accordingly, a Bayesian agent that naively assumes the beliefs of each of her neighbors were formed by a private observation (and not through repeated interaction with others) updates her beliefs proportionally to the product of her neighboring beliefs and likelihoods of her private signal [36]. In principle, the BWR updates replicate the rule that maps the initial priors, neighboring beliefs, and the private signal to the Bayesian belief at one time step for all future time steps [37]. Naivety of agents in these cases impedes their ability to learn; except in simple social structures such as cycles or rooted trees [38]. In [39] the authors show that learning in social networks with complex neighborhood structures can be achieved if agents choose a neighbor randomly at every round and restrict their belief update to the selected neighbor each time (essentially replicating the case of a directed circle where every neighborhood is a singlton). Geometric averaging and logarithmic opinion pools have a long history in Bayesian analysis and behavioral decision models [40], [41] and they can be also justified under specific behavioral assumptions [42]. The are also quite popular as a non-Bayesian update rule in distributed detection and estimation litrature [43], [44], [10], [14], [45]. In [45] the authors use a logarithmic opinion pool to combine the estimated posterior probability distributions in a Bayesian consensus filter; and show that as a result: the sum of KullbackLeibler divergences between the consensual probability distribution and the local posterior probability distributions is minimized. Minimizing the sum of KullbackLeibler divergences as a way to globally aggregate locally measured probability distributions is proposed in [46], [47] where the corresponding minimizer is dubbed the KullbackLeibler average. Similar interpretations of the log-linear update are offered in [48] as a gradient step for minimizing either the KullbackLeibler distance to the true distribution, or in [11] as a posterior incorporation of the most recent observations, such that the sum of KullbackLeibler distance to the local priors is minimized; indeed, the Bayes' rule itself has a product form and the Bayesian posterior can be characterized as the solution of an optimization problem involving the KullbackLeibler divergence to the prior distribution and subjected to the observed data [49].

In this paper, we highlight some key differences between the linear and log-linear update rules in the way they accommodate new observations. The requirement of averaging over time for linear updates necessitates that new observations be discounted as $1/t$ with increasing time, which avoids fluctuations with new observations in the limit. The same principle governs the discounting or diminishing step sizes in the case of consensus+innovation algorithm [50], as well as other online learning methods [51]. However, in case of log-linear update rules no such discounting is necessary. Because the product-nature of such rules imply that as beliefs approach a point mass their multiplication with the product of likelihoods of new observations will have less and less effect. This in turn allows us to effectively accommodate the varying sizes of data sets at every time-period using log-linear update rules with fixed coefficients.

Another key difference between the linear and log-linear updates is that the weights in the former need to be adjusted for the initial sample sizes, whereas the latter require no such adjustment of weights. Accordingly, in the linear case an agent weighs each neighbor's report differently and in accordance with the sample qualities or qualities (see (4)). On the other hand, when communicating their beliefs for log-linear updating the quality of each neighbor's signal is already internalized in their reported beliefs; hence, when incorporating its neighboring beliefs, an agent regards the reported beliefs of all its neighbors equally, and irrespective of the quality of their sample points. These observations lead to the conclusion that: log-*linear aggregation schemes (as opposed to linear ones) are very effective design tools for dealing with various types of heterogenities that arise in networked systems.*

## APPENDIX
## PROOFS OF THE MAIN RESULTS

### A. Proof of Theorem 1, Maximum Likelihood Estimation

We begin by forming the vectorized log-belief ratio updates as follows. Define

$$\phi_{i,t}(\hat{\theta}, \check{\theta}) := \log\left(\boldsymbol{\mu}_{i,t}(\hat{\theta})/\boldsymbol{\mu}_{i,t}(\check{\theta})\right),$$
$$\boldsymbol{\lambda}_i(\hat{\theta}, \check{\theta}) := \log\left(\boldsymbol{\gamma}_i(\hat{\theta})/\boldsymbol{\gamma}_i(\check{\theta})\right), \qquad (8)$$

and their vectorization

$$\overline{\boldsymbol{\phi}}_t(\hat{\theta}, \check{\theta}) := (\phi_{1,t}(\hat{\theta}, \check{\theta}), \ldots, \phi_{n,t}(\hat{\theta}, \check{\theta})),$$
$$\overline{\boldsymbol{\lambda}}(\hat{\theta}, \check{\theta}) := (\boldsymbol{\lambda}_1(\hat{\theta}, \check{\theta}), \ldots, \boldsymbol{\lambda}_n(\hat{\theta}, \check{\theta})). \qquad (9)$$

By forming the belief ratio $\boldsymbol{\mu}_{i,t}(\hat{\theta})/\boldsymbol{\mu}_{i,t}(\check{\theta})$, taking the logarithms of both sides, and using the vectorization in (9), we can rewrite the belief updates in (I) as a linear updated in terms of log ratios:

$$\overline{\boldsymbol{\phi}}_{t+1}(\hat{\theta}, \check{\theta}) = (I + A)\overline{\boldsymbol{\phi}}_t(\hat{\theta}, \check{\theta})$$
$$= (I + A)^{t+1}\overline{\boldsymbol{\phi}}_0(\hat{\theta}, \check{\theta}) = (I + A)^{t+1}\overline{\boldsymbol{\lambda}}(\hat{\theta}, \check{\theta}). \quad (10)$$

When the network graph $\mathcal{G}$ is connected, the matrix $I + A$ is primitive. The Perron-Frobenius theory [33, Theorems 1.5 and 1.7] implies that $I + A$ has a simple positive real eigenvalue equal to its spectral radius $\rho(I + A) = 2$. Moreover, the left and right eigenspaces associated with this eigenvalue are both one-dimensional and the corresponding eigenvectors can be taken to be both equal to $(1/\sqrt{n})\mathbb{1}$. The magnitude of any other eigenvalue of $I + A$ is strictly less than 2. Hence, the eigenvalues of $I + A$ denoted by $\alpha_i := \lambda_i(I + A)$, $i \in [n]$, which are all real, can be ordered as follows: $-2 < \lambda_n(I + A) \leq \lambda_{n-1}(I + A) \leq \ldots \leq \lambda_1(I + A) = 2$. Susequently, we can employ the eigendecomposition of $(I + A)$ to analyze the behavior of $(I + A)^{t+1}$ in (10). Specifically, we can take a set of bi-orthonormal vectors $\bar{l}_i, \bar{r}_i$ as the left and right eigenvectors corresponding to the $i$th eigenvalue of $I+A$, satisfying: $\|\bar{l}_i\|_2 = \|\bar{r}_i\|_2 = 1$, $\bar{l}_i^T \bar{r}_i = 1$ for all $i$ and $\bar{l}_i^T \bar{r}_j = 0$, $i \neq j$; in particular, $\bar{l}_1 = \bar{r}_1 = (1/\sqrt{n})\mathbb{1}$. Moreover, we have that [52]:

$$(I + A)^t = 2^t \left( \frac{1}{n}\mathbb{1}\mathbb{1}^T + \sum_{i=2}^n (\alpha_i/2)^t \bar{r}_i \bar{l}_i^T \right). \quad (11)$$

To proceed denote $\mathbf{\Lambda}(\hat{\theta}, \check{\theta}) := \mathbf{\Lambda}(\hat{\theta}) - \mathbf{\Lambda}(\check{\theta})$ and note that $\Lambda(\hat{\theta}, \check{\theta}) = \mathbb{1}^T \overline{\boldsymbol{\lambda}}(\hat{\theta}, \check{\theta})$. We can use (11) and (10), together with the fact that $|\alpha_i| < 2$ for all $i > 1$, established above using the Perron-Frobenius theory, to conclude that $\overline{\boldsymbol{\phi}}_t(\hat{\theta}, \check{\theta}) \to (2^t/n)\mathbb{1}\Lambda(\hat{\theta}, \check{\theta})$ almost surely. Moreover, since $\Theta^\star$ consists of the set of all maximizers of $\Lambda(\hat{\theta})$, we have that $\Lambda(\hat{\theta}, \tilde{\theta}) < 0$ whenever $\tilde{\theta} \in \Theta^\star$ and $\hat{\theta} \notin \Theta^\star$. Hence, for all $\tilde{\theta} \in \Theta^\star$ and any $\hat{\theta}$, $\phi_{i,t}(\hat{\theta}, \tilde{\theta}) \to -\infty$ if $\hat{\theta} \notin \Theta^\star$ and $\phi_{i,t}(\hat{\theta}, \tilde{\theta}) = 0$ whenever $\hat{\theta} \in \Theta^\star$; or equivalently, $\boldsymbol{\mu}_{i,t}(\hat{\theta})/\boldsymbol{\mu}_{i,t}(\tilde{\theta}) \to 0$ for all $\hat{\theta} \notin \Theta^\star$, while $\boldsymbol{\mu}_{i,t}(\hat{\theta}) = \boldsymbol{\mu}_{i,t}(\tilde{\theta})$ for any $\hat{\theta} \in \Theta^\star$. The latter together with the fact that $\sum_{\tilde{\theta} \in \Theta} \boldsymbol{\mu}_{i,t}(\tilde{\theta}) = 1$ for all $t$ implies that with probability one: $\lim_{t\to\infty} \boldsymbol{\mu}_{i,t}(\tilde{\theta}) = 1/|\Theta^\star|, \forall \tilde{\theta} \in \Theta^\star$ and $\lim_{t\to\infty} \boldsymbol{\mu}_{i,t}(\hat{\theta}) = 0, \forall \hat{\theta} \notin \Theta^\star$ as claimed.

Furthermore, we can use (11) and (10) to bound the distance between $\phi_{i,t}(\hat{\theta}, \check{\theta})$ and $(2^t/n)\Lambda(\hat{\theta}, \check{\theta})$ for any $i$, as follows:

$$\left| \phi_{i,t}(\hat{\theta}, \check{\theta}) - \frac{2^t}{n}\Lambda(\hat{\theta}, \check{\theta}) \right| \leq \left\| \overline{\boldsymbol{\phi}}_t(\hat{\theta}, \check{\theta}) - \frac{2^t}{n}\Lambda(\hat{\theta}, \check{\theta})\mathbb{1} \right\|_2$$

$$= \left\| \sum_{i=2}^n \left( \frac{\alpha_i}{2} \right)^t \bar{l}_i \bar{r}_i^T \overline{\boldsymbol{\lambda}}(\hat{\theta}, \check{\theta}) \right\|_2 \leq \sum_{i=2}^n \left| \frac{\alpha_i}{2} \right|^t \left| \bar{r}_i^T \overline{\boldsymbol{\lambda}}(\hat{\theta}, \check{\theta}) \right| \|\bar{l}_i\|_2$$

$$\leq \sum_{i=2}^n \left| \frac{\alpha_i}{2} \right|^t \left\| \overline{\boldsymbol{\lambda}}(\hat{\theta}, \check{\theta}) \right\|_2 \|\bar{r}_i\|_2 \|\bar{l}_i\|_2. \quad (12)$$

Orthonormality of the eigenvectors yields that $\|\bar{r}_i\|_2 = \|\bar{l}_i\|_2 = 1$; also by monotonicity of the $\ell_p$ norm we get that

$$\left\| \overline{\boldsymbol{\lambda}}(\hat{\theta}, \check{\theta}) \right\|_2 \leq \left\| \overline{\boldsymbol{\lambda}}(\hat{\theta}, \check{\theta}) \right\|_1 = \sum_{i=1}^n |\lambda_i(\hat{\theta}, \check{\theta})| \leq 2n\mathbf{L}_n,$$

where $\mathbf{L}_n = \max_{i \in [n]} \max_{\tilde{\theta} \in \Theta} |\log(\boldsymbol{\gamma}_i(\tilde{\theta}))|$ is the largest absolute log of product likelihoods that is achieved in the initial data sets, so that $|\boldsymbol{\lambda}_i(\hat{\theta}, \check{\theta})| = |\log(\boldsymbol{\gamma}_i(\hat{\theta})) - \log(\boldsymbol{\gamma}_i(\check{\theta}))| < 2\mathbf{L}_n$ for all $i$. Subsequently, (12) becomes

$$\left| \phi_{i,t}(\hat{\theta}, \check{\theta}) - \frac{2^t}{n}\Lambda(\hat{\theta}, \check{\theta}) \right| \leq 2\mathbf{L}_n n(n-1)(\alpha^\star)^t,$$

where $\alpha^\star = (1/2)\max\{\alpha_2, |\alpha_n|\}$ Hence,

$$\log(\boldsymbol{\mu}_{i,t}(\hat{\theta})) \leq \log\left( \frac{\boldsymbol{\mu}_{i,t}(\hat{\theta})}{\boldsymbol{\mu}_{i,t}(\tilde{\theta})} \right) = \phi_{i,t}(\hat{\theta}, \tilde{\theta})$$

$$\leq \frac{2^t}{n}\Lambda(\hat{\theta}, \tilde{\theta}) + 2\mathbf{L}_n n(n-1)(\alpha^\star)^t,$$

for all $i$ and any $\tilde{\theta} \in \Theta^\star$. Next suppose that the maximum likelihood estimator is unique so that $\Theta^\star = \{\theta^\star\}$ and let $\mathbf{l}_\Theta = \min_{\hat{\theta} \neq \theta^\star} \left| \Lambda(\hat{\theta}, \theta^\star) \right|$. Then for any $\hat{\theta} \neq \theta^\star$ and all agents $i$ we can bound the belief on $\hat{\theta}$ as follows:

$$\log(\boldsymbol{\mu}_{i,t}(\hat{\theta})) \leq -\frac{2^t}{n}\mathbf{l}_n + 2\mathbf{L}_n n(n-1)(\alpha^\star)^t.$$

Therefore, if we take

$$\mathbf{T} = \max\left\{ 1 + \frac{\log\left( \frac{n\log(n-1)}{\mathbf{l}_n} \right)}{\log 2}, \frac{\log\left( \frac{\log(n-1)}{2\mathbf{L}_n n(n-1)} \right)}{\log(\alpha^\star)} \right\}, \quad (13)$$

then for all $t > \mathbf{T}$, $\log(\boldsymbol{\mu}_{i,t}(\hat{\theta})) \leq -\log(n-1)$ so that $\boldsymbol{\mu}_{i,t}(\hat{\theta}) < \frac{1}{n-1} < \boldsymbol{\mu}_{i,t}(\theta^\star)$ for all $\hat{\theta} \neq \theta^\star$ and any $i \in [n]$; whence, any agent $i \in [n]$ can recover $\theta^\star$ as $\theta^\star = \arg\max_{\tilde{\theta} \in \Theta} \boldsymbol{\mu}_{i,t}(\tilde{\theta})$ at all $t > \mathbf{T}$. $\square$

*B. Proof of Theorem 2, Learning from Intermittent Streams*

The belief update rule proposed in (II) is the same as the time-invariant log-linear update with weighted self-beliefs considered in [35, Equation (13)]; except that here at every round each agent is receiving a random number of signals. Hence, the proof of convergence in [35, Equation (13)] can be applied here and with minor modifications. Specifically, we let $\overline{\boldsymbol{\phi}}_t(\hat{\theta}, \check{\theta})$ be the vectorized log belief ratio statistics as defined in (8) and (9), and define the log ratio of the likelihood products of the received signals: $\boldsymbol{\lambda}_{i,t}(\hat{\theta}, \check{\theta}) = \log(\boldsymbol{\gamma}_{i,t}(\hat{\theta})/\boldsymbol{\gamma}_{i,t}(\check{\theta}))$, and its vectorization $\overline{\boldsymbol{\lambda}}_t(\hat{\theta}, \check{\theta}) = (\boldsymbol{\lambda}_{1,t}(\hat{\theta}, \check{\theta}), \ldots, \boldsymbol{\lambda}_{n,t}(\hat{\theta}, \check{\theta}))$. Then after forming the log belief ratios, (II) in vectorized form yields that: $\overline{\boldsymbol{\phi}}_t(\hat{\theta}, \check{\theta}) = A\overline{\boldsymbol{\phi}}_{t-1}(\hat{\theta}, \check{\theta}) + \overline{\boldsymbol{\lambda}}_t(\hat{\theta}, \check{\theta}) = \sum_{\tau=0}^t A^\tau \overline{\boldsymbol{\lambda}}_{t-\tau}(\hat{\theta}, \check{\theta})$ and the latter converges almost surely to $\left((t/n)\mathbb{1}^T \mathbb{E}\{\overline{\boldsymbol{\lambda}}_0(\hat{\theta}, \check{\theta})\}\right)\mathbb{1}$, as $t \to \infty$; this is a simple consequecen of the Cesàro mean together with the strong law of large numbers. The proof follows since $\mathbb{E}\{\overline{\boldsymbol{\lambda}}_0(\hat{\theta}, \check{\theta})\} = (\Lambda_1(\hat{\theta}, \check{\theta}), \ldots, \Lambda_n(\hat{\theta}, \check{\theta}))^T$; in particular, $\lim_{t\to\infty} \frac{1}{t}\overline{\boldsymbol{\phi}}_t(\hat{\theta}, \check{\theta}) = \left((1/n)\sum_{i=1}^n \Lambda_i(\hat{\theta}, \check{\theta})\right)\mathbb{1}$, with probability one and whenever $\sum_{i=1}^n \Lambda_i(\hat{\theta}, \check{\theta}) < 0$, the agents learn the truth asymptotically exponentially fast, at the rate $\min_{\hat{\theta} \neq \theta} \left\{ (-1/n)\sum_{i=1}^n \Lambda_i(\hat{\theta}, \theta) \right\}$. $\square$

*C. Proof of Theorem 3, Minimum Variance Unbiased Estimation*

Define the concatenated variables $\overline{\boldsymbol{\mu}}_t = (\boldsymbol{\mu}_{1,t}, \ldots, \boldsymbol{\mu}_{n,t})^T$, $\overline{\boldsymbol{\lambda}} = (\xi(\mathbf{s}_1), \ldots, \xi(\mathbf{s}_n))^T$ and note that $\mathbf{m}_n = (1/n)\sum_{i=1}^n \xi(\mathbf{s}_i) = (1/n)\mathbb{1}^T \overline{\boldsymbol{\lambda}}$. Initialized by $\overline{\boldsymbol{\mu}}_0 = \overline{\boldsymbol{\lambda}}$, the evolution of beliefs under (III) can be written in the following vectorized form: $\overline{\boldsymbol{\mu}}_t = A\overline{\boldsymbol{\mu}}_{t-1} = A^t\overline{\boldsymbol{\lambda}}$, and as in Appendix A for a connected network $\mathcal{G}$, we have that $\lim_{t\to\infty} A^t = (1/n)\mathbb{1}\mathbb{1}^T$; and subsequently, $\lim_{t\to\infty} \boldsymbol{\mu}_t = (1/n)\mathbb{1}\mathbb{1}^T\overline{\boldsymbol{\lambda}} = \mathbb{1}\mathbf{m}_n$. Hence, the claim about the almost sure limits of every agents' beliefs is verified. To investigate the rate of convergence of $\overline{\boldsymbol{\mu}}_t$ to $\mathbb{1}\mathbf{m}_n$ we can write: $\overline{\boldsymbol{\mu}}_t - \mathbb{1}\mathbf{m}_n = (A^t - (1/n)\mathbb{1}\mathbb{1}^T)\overline{\boldsymbol{\lambda}}$. Hence for each node $i$ we have:

$$|\boldsymbol{\mu}_{i,t} - \mathbf{m}_n| = \left|\sum_{j=1}^n \left([A^t]_{ij} - \frac{1}{n}\right)\xi(\mathbf{s}_j)\right|$$

$$\leq \sum_{j=1}^n \left|[A^t]_{ij} - \frac{1}{n}\right||\xi(\mathbf{s}_j)| \leq \mathbf{M}_n \sum_{j=1}^n \left|[A^t]_{ij} - \frac{1}{n}\right|. \quad (14)$$

We next use the fact that $A$ can specify the transition probabilities of an aperiodic irreducible Markov chain with uniform stationary distribution. In particular, it is a time-reversible Markov chain and [53, Proposition 3] implies that

$$\sum_{j=1}^n \left|[A^t]_{ij} - \frac{1}{n}\right| \leq \sqrt{(n-1)}\,(\beta^\star)^t, \quad (15)$$

where $\beta^\star = \max\{\lambda_2(A), |\lambda_n(A)|\}$ and $0 \leq \beta^\star < 1$ as a consequence of Perron-Frobenius theory [33, Theorems 1.5 and 1.7] applied to the primitive matrix $A$. Replacing (15) in (14) yields that for all $i$ the distance to the limiting values $\mathbf{m}_n$ decrease at least exponential fast and can be bounded as follows: $|\boldsymbol{\mu}_{i,t} - \mathbf{m}_n| \leq \mathbf{M}_n\sqrt{(n-1)}\,(\beta^\star)^t$. The claimed finite time guarantee now follows upon setting $\mathbf{M}_n\sqrt{(n-1)}\,(\beta^\star)^t < \epsilon$ or equivalently: $t > \left(\log(\epsilon) - \log(\mathbf{M}_n\sqrt{n-1})\right)/\log\beta^\star$. □

### D. Proof of Theorem 4, Online Learning of Expected Values

Let $\overline{\boldsymbol{\mu}}_t$ be as in the proof of Theorem 3 (Appendix C), $\overline{\boldsymbol{\lambda}}_t = (\xi(\mathbf{s}_{1,t}), \ldots, \xi(\mathbf{s}_{n,t}))^T$, and $A_t = \frac{t-1}{t}A$. Under (IV) the beliefs evolve as follows:

$$\overline{\boldsymbol{\mu}}_t = A_t\overline{\boldsymbol{\mu}}_{t-1} + \frac{1}{t}\overline{\boldsymbol{\lambda}}_t = \frac{1}{t}\overline{\boldsymbol{\lambda}}_t + \sum_{\tau=1}^{t-1}\left(\prod_{u=\tau+1}^t A_u\right)\frac{1}{\tau}\overline{\boldsymbol{\lambda}}_\tau$$

$$= \frac{1}{t}\overline{\boldsymbol{\lambda}}_t + \sum_{\tau=1}^{t-1}\left(\frac{t-1}{t}\times\frac{t-2}{t-1}\times\ldots\times\frac{\tau}{\tau+1}A^{t-\tau}\right)\frac{1}{\tau}\overline{\boldsymbol{\lambda}}_\tau$$

$$= \frac{1}{t}\sum_{\tau=1}^t A^{t-\tau}\overline{\boldsymbol{\lambda}}_\tau. \quad (16)$$

As in Appendices B and C, we have that $\lim_{\tau\to\infty} A^\tau = (1/n)\mathbb{1}\mathbb{1}^T$, and we can invoke the Cesàro mean together with the strong law to conclude that

$$\lim_{t\to\infty}\overline{\boldsymbol{\mu}}_t = \mathbb{1}\left(\lim_{t\to\infty}\frac{1}{nt}\sum_{\tau=1}^t\sum_{i=1}^n \xi(\mathbf{s}_{i,\tau})\right) = \mathbb{1}\mathbb{E}\{\xi(\mathbf{s}_{i,1})\},$$

so that $\boldsymbol{\mu}_{i,t} \to m_\theta$ with probability one for all agents $i \in [n]$; in particular, $\boldsymbol{\mu}_{i,t}$ for each $i$ is a strongly consistent estimator of $\overline{\theta}$. We can further bound the rate of decrease in $\text{Var}(\mu_{i,t})$ as $t$ increases. Taking expectation of both sides in (16) yields that $\mathbb{E}\{\overline{\boldsymbol{\mu}}_t\} = (1/t)\sum_{\tau=1}^t A^{t-\tau}\mathbb{1}m_\theta = \mathbb{1}m_\theta$. Hence, we can subtract $\mathbb{1}m_\theta$ from both sides of (16) and bound the variance of $\overline{\boldsymbol{\mu}}_t$ in terms of the variance of i.i.d. random variable $\overline{\boldsymbol{\lambda}}_t$ and rate of convergence (mixing) for $A^t \to (1/n)\mathbb{1}\mathbb{1}^T$. Indeed, using (16) we can write

$$|\boldsymbol{\mu}_{i,t} - m_\theta| = \left|\frac{1}{t}\sum_{\tau=1}^t\sum_{j=1}^n [A^{t-\tau}]_{ij}\xi(\mathbf{s}_{j,\tau}) - m_\theta\right|.$$

Next by adding and subtracting $\frac{1}{nt}\sum_{\tau=1}^t\sum_{j=1}^n \xi(\mathbf{s}_{j,\tau})$, which is the average of all signals across all times and agents; and then applying the triangle inequality we obtain:

$$|\boldsymbol{\mu}_{i,t} - m_\theta| \leq \frac{1}{t}\left|\sum_{\tau=1}^t\sum_{j=1}^n\left([A^{t-\tau}]_{ij} - \frac{1}{n}\right)\xi(\mathbf{s}_{j,\tau})\right|$$

$$+ \frac{1}{nt}\left|\sum_{\tau=1}^t\sum_{j=1}^n (\xi(\mathbf{s}_{j,\tau}) - m_\theta)\right|.$$

Taking squares of both sides and using the Cauchy-Schwartz inequality yields that

$$(\boldsymbol{\mu}_{i,t} - m_\theta)^2 \leq \qquad (17)$$

$$\frac{1}{t^2}\left\{\sum_{\tau=1}^t\sum_{j=1}^n\left([A^{t-\tau}]_{ij} - \frac{1}{n}\right)^2\right\}\left\{\sum_{\tau=1}^t\sum_{j=1}^n \xi(\mathbf{s}_{j,\tau})^2\right\}$$

$$+ \frac{1}{n^2t^2}\left(\sum_{\tau=1}^t\sum_{j=1}^n (\xi(\mathbf{s}_{j,\tau}) - m_\theta)\right)^2.$$

We next apply the Markov chain mixing time inequality (15) from the proof of Theorem 3 (Appendix C) to bound

$$\sum_{\tau=1}^t\sum_{j=1}^n \left([A^{t-\tau}]_{ij} - \frac{1}{n}\right)^2 \leq \sum_{\tau=1}^t \left(\sum_{j=1}^n \left|[A^{t-\tau}]_{ij} - \frac{1}{n}\right|\right)^2$$

$$\leq \sum_{\tau=1}^t (n-1)(\beta^\star)^{2(t-\tau)} \leq \frac{n-1}{1-\beta^{\star 2}}, \quad (18)$$

where $\beta^\star = \max\{\lambda_2(A), |\lambda_n(A)|\}$ and $0 \leq \beta^\star < 1$. Furthermore, since $\{\xi(\mathbf{s}_{j,\tau}), j \in [n], t \in \mathbb{N}\}$ form a sequence of i.i.d. random variables with mean $m_\theta$, we have:

$$\mathbb{E}\left\{\left(\sum_{\tau=1}^t\sum_{j=1}^n (\xi(\mathbf{s}_{j,\tau}) - m_\theta)\right)^2\right\} = nt\text{Var}\{\xi(\mathbf{s}_{j,\tau})\}.$$

We can now bound $\text{Var}\{\boldsymbol{\mu}_{i,t}\}$ by taking expectations of both sides in (17) and using (18) to get (5) and subsequently (6) in Section IV; whence, $\text{Var}\{\boldsymbol{\mu}_{i,t}\} = O(1/t)$ as claimed. □

## REFERENCES

[1] V. Borkar and P. Varaiya, "Asymptotic agreement in distributed estimation," *IEEE Transactions on Automatic Control*, vol. 27, no. 3, pp. 650–655, Jun 1982.

[2] J. N. Tsitsiklis and M. Athans, "Convergence and asymptotic agreement in distributed decision problems," *Automatic Control, IEEE Transactions on*, vol. 29, no. 1, pp. 42–50, 1984.

[3] J. N. Tsitsiklis, "Decentralized detection," *Advances in Statistical Signal Processing*, vol. 2, no. 2, pp. 297–344, 1993.

[4] J.-F. Chamberland and V. V. Veeravalli, "Decentralized detection in sensor networks," *IEEE Transactions on Signal Processing*, vol. 51, no. 2, pp. 407–416, 2003.

[5] R. Olfati-Saber and J. Shamma, "Consensus filters for sensor networks and distributed sensor fusion," *IEEE Conference on Decision and Control*, pp. 6698 – 6703, 2005.

[6] S. Kar, J. Moura, and K. Ramanan, "Distributed parameter estimation in sensor networks: Nonlinear observation models and imperfect communication," *IEEE Transactions on Information Theory*, vol. 58, no. 6, pp. 3575–3605, 2012.

[7] N. A. Atanasov, J. Le Ny, and G. J. Pappas, "Distributed algorithms for stochastic source seeking with mobile robot networks," *Journal of Dynamic Systems, Measurement, and Control*, 2014.

[8] N. Atanasov, R. Tron, V. M. Preciado, and G. J. Pappas, "Joint estimation and localization in sensor networks," *IEEE Conference on Decision and Control (CDC)*, pp. 6875–6882, 2014.

[9] S. Shahrampour, A. Rakhlin, and A. Jadbabaie, "Distributed detection: Finite-time analysis and impact of network topology," *IEEE Transactions on Automatic Control*, 2016.

[10] A. Nedić, A. Olshevsky, and C. A. Uribe, "Nonasymptotic convergence rates for cooperative learning over time-varying directed graphs," *American Control Conference (ACC)*, pp. 5884–5889, 2015.

[11] ——, "Fast convergence rates for distributed non-bayesian learning," *arXiv preprint arXiv:1508.05161*, 2015.

[12] V. Krishnamurthy and H. V. Poor, "Social learning and bayesian games in multiagent signal processing: How do local and global decision makers interact?" *IEEE Signal Processing Magazine,*, vol. 30, no. 3, pp. 43–57, 2013.

[13] Y. Wang and P. M. Djuric, "Social learning with bayesian agents and random decision making," *IEEE Transactions on Signal Processing*, vol. 63, no. 12, pp. 3241–3250, 2015.

[14] A. Lalitha, A. Sarwate, and T. Javidi, "Social learning and distributed hypothesis testing," *IEEE International Symposium on Information Theory*, pp. 551–555, 2014.

[15] M. H. DeGroot, "Reaching a consensus," *Journal of American Statistical Association*, vol. 69, pp. 118 – 121, 1974.

[16] A. Jadbabaie, P. Molavi, A. Sandroni, and A. Tahbaz-Salehi, "Non-bayesian social learning," *Games and Economic Behavior*, vol. 76, no. 1, pp. 210 – 225, 2012.

[17] L. Xiao, S. Boyd, and S. Lall, "A scheme for robust distributed sensor fusion based on average consensus," in *Fourth International Symposium on Information Processing in Sensor Networks (IPSN)*, 2005, pp. 63–70.

[18] ——, "A space-time diffusion scheme for peer-to-peer least-squares estimation," in *Proceedings of the 5th international conference on Information Processing in Sensor Networks*, 2006, pp. 168–176.

[19] R. J. Aumann, "Agreeing to disagree," *The annals of statistics*, pp. 1236–1239, 1976.

[20] J. D. Geanakoplos and H. M. Polemarchakis, "We can't disagree forever," *Journal of Economic Theory*, vol. 28, no. 1, pp. 192–200, 1982.

[21] A. D. Dimakis, A. D. Sarwate, and M. J. Wainwright, "Geographic gossip: Efficient averaging for sensor networks," *Signal Processing, IEEE Transactions on*, vol. 56, no. 3, pp. 1205–1216, 2008.

[22] A. Jadbabaie, J. Lin, and A. S. Morse, "Coordination of groups of mobile autonomous agents using nearest neighbor rules," *IEEE Transactions on Automatic Control*, vol. 48, no. 6, pp. 988–1001, 2003.

[23] M. Mesbahi and M. Egerstedt, *Graph Theoretic Methods in Multiagent Networks*. Princeton University Press, 2010.

[24] F. Bullo, J. Cortés, and S. Martínez, *Distributed Control of Robotic Networks*. Princeton University Press, 2009.

[25] S. Boyd, P. Diaconis, and L. Xiao, "Fastest mixing markov chain on a graph," *SIAM review*, vol. 46, no. 4, pp. 667–689, 2004.

[26] M. A. Rahimian and A. Jadbabaie, "Group decision making and social learning," *55th IEEE Conference on Decision and Control (CDC)*, 2016.

[27] ——, "Bayesian heuristics for group decisions," 2016, preprint.

[28] T. M. Cover and J. A. Thomas, *Elements of Information Theory*, ser. A Wiley-Interscience publication. Wiley, 2006.

[29] P. J. Bickel and K. A. Doksum, *Mathematical Statistics: Basic Ideas and Selected Topics, volume I*. CRC Press, 2015.

[30] G. Casella and R. L. Berger, *Statistical inference*. Duxbury Pacific Grove, CA, 2002, vol. 2.

[31] A. Olshevsky, "Linear time average consensus on fixed graphs and implications for decentralized optimization and multi-agent control," *arXiv preprint arXiv:1411.4186*, 2014.

[32] D. A. Levin, Y. Peres, and E. L. Wilmer, *Markov Chains and Mixing Times*. American Mathematical Society, 2009.

[33] E. Seneta, *Non-negative matrices and Markov chains*. Springer, 2006.

[34] L. J. Billera and P. Diaconis, "A geometric interpretation of the metropolis-hastings algorithm," *Statistical Science*, pp. 335–339, 2001.

[35] M. A. Rahimian and A. Jadbabaie, "Learning without recall: A case for log-linear learning," *5th IFAC Workshop on Distributed Estimation and Control in Networked Systems*, 2015.

[36] M. A. Rahimian, P. Molavi, and A. Jadbabaie, "(Non-) bayesian learning without recall," *IEEE Conference on Decision and Control (CDC)*, pp. 5730–5735, 2014.

[37] M. A. Rahimian and A. Jadbabaie, "Bayesian learning without recall," *arXiv preprint arXiv:1601.06103*, 2016.

[38] ——, "Learning without recall in directed circles and rooted trees," *American Control Conference*, pp. 4222–4227, 2015.

[39] M. A. Rahimian, S. Shahrampour, and A. Jadbabaie, "Learning without recall by random walks on directed graphs," *IEEE Conference on Decision and Control (CDC)*, 2015.

[40] G. L. Gilardoni and M. K. Clayton, "On reaching a consensus using DeGroot's iterative pooling," *The Annals of Statistics*, pp. 391–401, 1993.

[41] M. J. Rufo, J. Martin, C. J. Pérez *et al.*, "Log-linear pool to combine prior distributions: A suggestion for a calibration-based approach," *Bayesian Analysis*, vol. 7, no. 2, pp. 411–438, 2012.

[42] P. Molavi, A. Tahbaz-Salehi, and A. Jadbabaie, "Foundations of non-bayesian social learning," *Columbia Business School Research Paper*, 2015.

[43] S. Shahrampour and A. Jadbabaie, "Exponentially fast parameter estimation in networks using distributed dual averaging," *52nd IEEE Conference on Decision and Control (CDC)*, pp. 6196–6201, 2013.

[44] K. Rahnama Rad and A. Tahbaz-Salehi, "Distributed parameter estimation in networks," *49th IEEE Conference on Decision and Control (CDC)*, pp. 5050–5055, 2010.

[45] S. Bandyopadhyay and S.-J. Chung, "Distributed estimation using bayesian consensus filtering," *American Control Conference*, pp. 634–641, 2014.

[46] G. Battistelli, L. Chisci, S. Morrocchi, and F. Papi, "An information-theoretic approach to distributed state estimation," *IFAC Proceedings Volumes*, vol. 44, no. 1, pp. 12 477–12 482, 2011.

[47] G. Battistelli and L. Chisci, "Kullback–leibler average, consensus on probability densities, and distributed state estimation with guaranteed stability," *Automatica*, vol. 50, no. 3, pp. 707–718, 2014.

[48] A. Nedić, A. Olshevsky, and C. Uribe, "Distributed learning with infinitely many hypotheses," *arXiv preprint arXiv:1605.02105*, 2016.

[49] A. Zellner, "Optimal information processing and bayes's theorem," *The American Statistician*, vol. 42, no. 4, pp. 278–280, 1988.

[50] S. Kar and J. Moura, "Consensus+ innovations distributed inference over networks: cooperation and sensing in networked systems," *Signal Processing Magazine, IEEE*, vol. 30, no. 3, pp. 99–109, 2013.

[51] N. E. Leonard and A. Olshevsky, "Cooperative learning in multiagent systems from intermittent measurements," *SIAM Journal on Control and Optimization*, vol. 53, no. 1, pp. 1–29, 2015.

[52] G. James and V. Rumchev, "Stability of positive linear discrete-time systems," *Bulletin of the Polish Academy of Sciences. Technical Sciences*, vol. 53, no. 1, pp. 1–8, 2005.

[53] P. Diaconis and D. Stroock, "Geometric bounds for eigenvalues of markov chains," *The Annals of Applied Probability*, pp. 36–61, 1991.